\documentclass[aps,prb,superscriptaddress,twocolumn,showpacs]{revtex4-2}
\usepackage{amssymb}
\usepackage{epsfig}
\DeclareMathAlphabet{\pazocal}{OMS}{zplm}{m}{n}            
\usepackage{graphicx}
\usepackage{color}
\usepackage{bm}
\usepackage[normalem]{ulem}     
\usepackage{soul}
\usepackage{amsmath}
\usepackage{braket} 
\usepackage{rotating}
\usepackage{multirow} 
\DeclareMathAlphabet{\pazocal}{OMS}{zplm}{m}{n}            
\usepackage{graphicx}
\usepackage{hyperref}
\usepackage{mhchem}

\begin{document}
\title{Non-relativistic spin splitting: Features and Functionalities}

\author{Sayantika Bhowal}
\email{sbhowal@iitb.ac.in}
\affiliation{Department of Physics, Indian Institute of Technology Bombay, Mumbai 400076, India}

\author{Arnab Bose}
\affiliation{Department of Electrical Engineering, Indian Institute of Technology Kanpur, Kanpur 208016, India}

\date{\today}

\begin{abstract}
Recently, spin splitting of non-relativistic origin in compensated antiferromagnets has drawn growing attention in condensed matter research. Although many materials, now known to exhibit such spin splitting, have been studied for decades, their manifestation along non-high-symmetry momentum directions initially hindered their recognition. In recent years, significant progress has been made in uncovering the symmetry principles that allow non-relativistic spin splitting in the absence of net magnetization, revealing the unconventional physics arising from their coexistence. In this review, we provide a concise overview of non-relativistic spin splitting in compensated antiferromagnets with various spin configurations, including collinear, coplanar, and non-coplanar spin arrangements. We summarize practical identification guidelines, highlight characteristic features in electronic band structures, and discuss the emerging functionalities, with an emphasis on promising directions for future exploration.   

\end{abstract}

\maketitle

\section{Introduction}

\begin{figure*}[t]
    \centering
    \includegraphics[width=\textwidth]{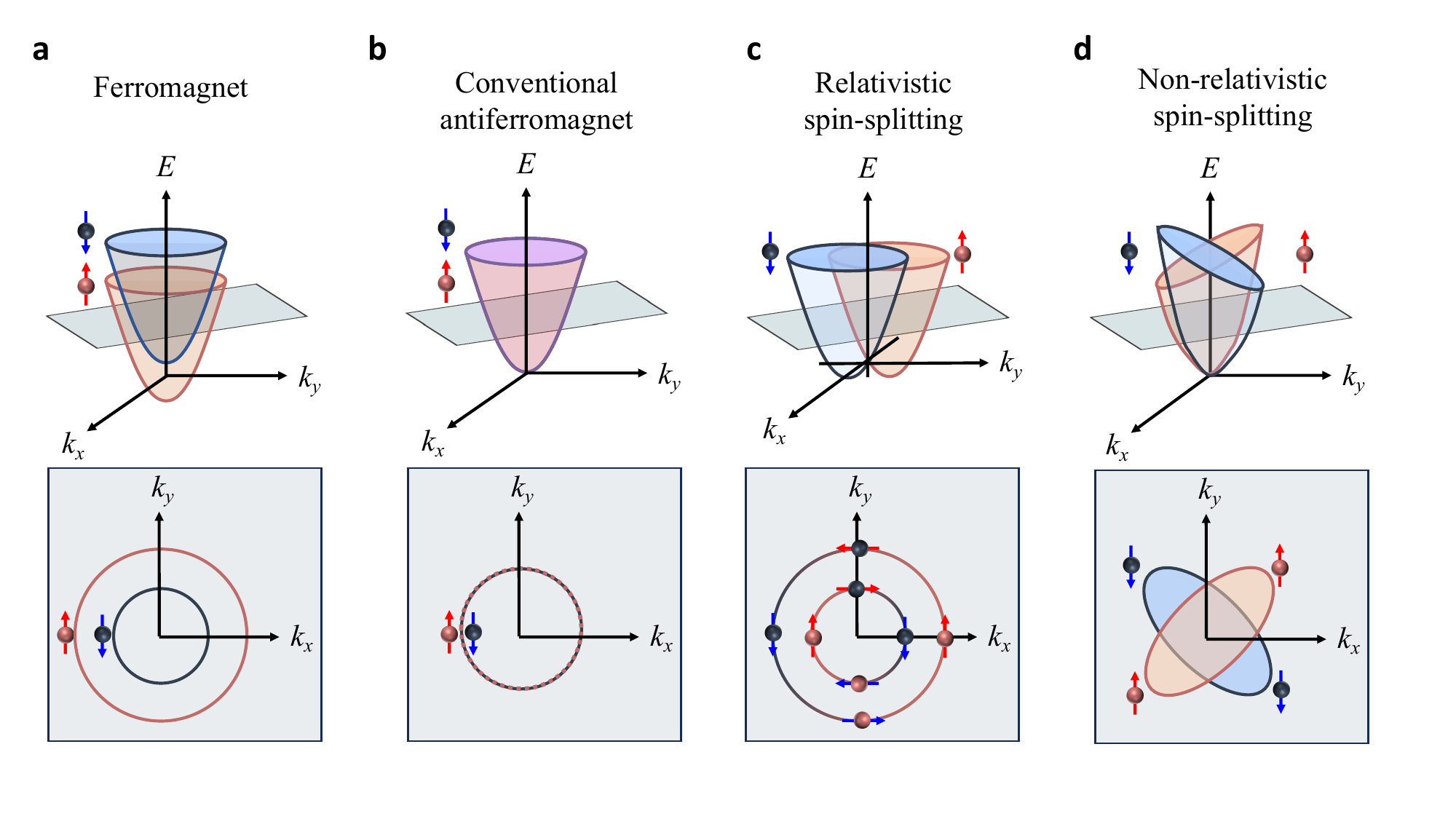}
    \caption{Schematic illustration of (a) spin splitting in ferromagnets due to exchange splitting, (b) spin-degenerate bands in conventional $\mathcal{IT}$-symmetric AFMs, (c) relativistic Rashba-like spin splitting, and (d) NRSS arising from broken $\mathcal{IT}$ symmetry in the absence of magnetization. The lower panel shows the corresponding energy cross-sections on the $k_x$–$k_y$ plane, as indicated by the shaded plane in the upper panel.}
    \label{fig1}
\end{figure*}

Moore’s prediction of transistor doubling every two years \cite{Moore1965} has long driven semiconductor innovation. As conventional technologies near physical limits, spintronics, which encodes information using the spin moment (``up" or ``down") of an electron, offers a promising path forward \cite{Igor2004,Hirohata2020,Guo2024,Khalili2024}. While traditional spintronic devices rely on ferromagnets with exchange-split spin-polarized bands (see Fig. \ref{fig1}a), their sensitivity to stray magnetic fields poses limitations. In contrast, antiferromagnetic spintronics \cite{Baltz2018,Fukami2020,Dal2024} offers robustness against such fields. However, antiferromagnets (AFMs) were traditionally thought to have degenerate spin-polarized bands as shown in Fig. \ref{fig1}b, limiting their usefulness. Energy splitting between spin-polarized bands in the reciprocal space, essential for spintronics, can emerge in AFMs and even nonmagnetic systems via spin-orbit coupling (SOC), as in the Rashba and Dresselhaus effects \cite{PekarRashba1964, Rashba1960, BychkovRashba,Winkler2003,Manchon2015,Koo2020} (see Fig. \ref{fig1}c), for example, but this typically restricts materials to heavy-element systems with strong SOC.

In this context, recent findings, revealing splitting between spin polarized bands of a compensated AFM, as depicted in Fig. \ref{fig1}d, in contrast to the conventional AFMs with degenerate up and down spin-polarized bands, are particularly intriguing, addressing the key limitations in spintronics research \cite{Hayami2019, Yuan2020, Yuan2021, Smejkal2022PRX, YuanZunger2023, Guo2023, Zeng2024, Lee2024PRL, Krempask2024, Reimers2024, Aoyama2024, Lin2024, Kyo-Hoon2019, Libor2020, Paul2024, Bai2024}. Remarkably, this splitting can occur even in the absence of SOC, distinguishing it from conventional SOC-driven Rashba and Dresselhaus effects. Consequently, the magnitude of NRSS is typically much larger than that of Rashba-like spin splitting, which is constrained by the relatively weak SOC strength. 

Going beyond spintronic applications, the emergence of NRSS without any net magnetization is also of fundamental interest, representing the long-sought $p$-wave magnetic analogue of superfluid $^3$He and the magnetic counterpart of $d$-wave superconductors \cite{Schofield2009,Libor2022PRX,Jungwirth2024,Hellenes2024}. By combining traits of ferromagnets and conventional AFMs, NRSS AFMs offer the platform for achieving otherwise difficult properties, such as efficient spin-current generation \cite{Naka2019, Hernandez2021, Shao2021, Bose2022,Hu2024}, spin-splitting torque \cite{Bai2022, Karube2021}, giant magnetoresistance \cite{Libor2022}, spontaneous Hall effect \cite{Libor2022NatRev, Helena2020, Feng2022, Betancourt2023, Cheong2024, Sato2024} without any magnetization as well as unconventional superconducting properties \cite{Mazin2022,Zhu2023,Banerjee2024,Chakraborty2024,Zhang2024,Lee2024}, enhanced thermal transport \cite{Zhou2024,Yershov2024}, magnetoelectric-multiferroic properties, piezomagnetic and kinetomagnetic effects.

Early developments in this field were focused on identifying the underlying symmetries \cite{Yuan2021, Smejkal2022PRX, Guo2023, Zeng2024}. To analyze the presence of spin splitting in the absence of SOC, the spin group theory, in which the symmetry operations on the spin-space and the structure act differently, was invoked. While initially, the inversion symmetric collinear AFMs were investigated for the NRSS \cite{Yuan2021, Smejkal2022PRX, Guo2023, Zeng2024}, later it was broadened to include also non-collinear spin systems and inversion-broken systems \cite{Hayami2020,Jungwirth2024,Hellenes2024,Radaelli2025}. Consequently, the presence of intriguing hidden magnetic order, characterizing the highly anisotropic magnetization density in these spin-split systems beyond the antiferromagnetic dipole order, has also been identified \cite{Hayami2019,HayamiPRB2020,BhowalSpaldin2022,Nag2024,Verbeek2024}. Guided by the established symmetry requirements, a wide range of previously known antiferromagnetic materials, both metals and insulators, have been identified as potential candidates for hosting NRSS. Interestingly, such materials are quite common in nature and have been extensively studied over the years. However, the NRSS feature has often been overlooked, as in most cases, the spin splitting does not occur along the high-symmetry $\vec{k}$ paths. 

Following the discovery of the NRSS in real materials, the current interest has shifted to controlling the NRSS \cite{Duan2025, Gu2025,Bandyopadhyay2024,Bandyopadhyay2025,Libor2025,Karetta2025}. Several interesting review articles \cite{Smejkal2022NatRev,Libor2022PRX, Bai2024,Tamang2024,Song2025,Jungwirth2024} have already appeared in this rapidly growing field. In the present review, we aim to provide a concise overview of different types of NRSS proposed so far, with a focus on their characteristic features, candidate materials, and an overview of some of the unconventional functionalities that they give rise to. We conclude by discussing potential design strategies for tuning this spin splitting and some promising future directions in the field.

\section{Classification and Characteristics}\label{sec2}

Over the past years, the concept of NRSS has broadened to include both collinear and non-collinear antiferromagnetic spin configurations, with or without inversion symmetry, to describe symmetric and antisymmetric spin splitting \cite{Hayami2020PRB,Jungwirth2024,Cheong2024,Radaelli2025}. Here, we summarize the various NRSS types arising from different symmetry conditions and their signatures in the band structure. Before exploring these classes, we highlight the key requirement for spin splitting based on how the band energy $\epsilon(\vec k, \uparrow)$ at momentum $\vec k$ with up spin polarization transforms under time-reversal ($\cal T$) and inversion ($\cal I$):
 \begin{eqnarray} \label{ITsym}\nonumber
   \cal T &:& \epsilon(\vec k, \uparrow) \rightarrow \epsilon(-\vec k, \downarrow)  \\ 
   \cal I &:& \epsilon(\vec k, \uparrow) \rightarrow \epsilon(-\vec k, \uparrow).
 \end{eqnarray}

\begin{figure*}[t]
    \centering
\includegraphics[width=\textwidth]{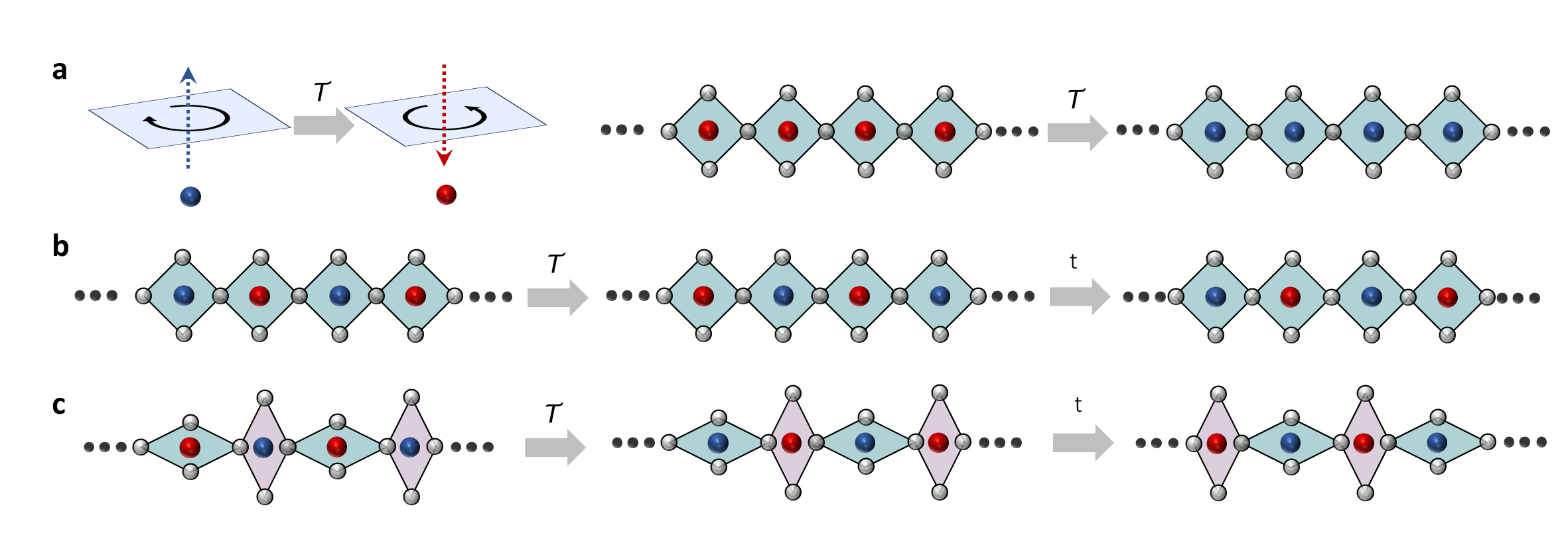}
    \caption{Schematic illustration of the effect of time-reversal $\cal T$ symmetry on (a) ferromagnets, (b) AFMs with equivalent surrounding nonmagnetic environments around the two magnetic sublattices, and (c) AFMs with inequivalent surrounding nonmagnetic environments around the two magnetic sublattices. While all ferromagnets break $\cal T$ symmetry due to the reversal of the spin moment direction, as shown in (a), AFMs may or may not preserve the global $\cal T$ symmetry, i.e., time-reversal plus translation $t$, as shown in (b) and (c), respectively. }
    \label{fig2}
\end{figure*}

Thus, the presence of combined $\cal{IT}$ symmetry implies $\epsilon(\vec k, \uparrow) \rightarrow \epsilon(\vec k, \downarrow)$, enforcing spin degeneracy (Kramers degeneracy) throughout the Brillouin zone (BZ). Thus, breaking $\cal{IT}$ is necessary for NRSS and can occur via three scenarios:  
(i) $\cal T$ broken, $\cal I$ preserved;  
(ii) $\cal I$ broken, $\cal T$ preserved;  
(iii) both $\cal T$ and $\cal I$ broken. 
It is important to point out that here, and throughout the rest of the manuscript, $\cal T$ refers to the global time-reversal symmetry (time-reversal combined with translation). As illustrated in Fig.~\ref{fig2}, AFMs may or may not preserve $\cal T$ symmetry, even though local $\cal T$ symmetry is always broken.
Also, while broken $\cal{IT}$ symmetry is a necessary condition for NRSS, it is not sufficient, as spin splitting can also result from SOC. This is where the spin group theory helps identify NRSS in the absence of SOC. While we refer readers to Refs. \cite{Libor2022,Liu2022,Xiao2024,Chen2024,Jiang2024,Watanabe2024} for a detailed and insightful treatment of spin group theory, here we focus instead on quick identification rules and the characteristic features in the band structure of the various NRSS types.

\subsection{$\cal T$ broken, $\cal I$ preserved} \label{case1}

\begin{itemize}

 {\item {\bf Colinear AFMs with symmetric spin splitting}-
 Initial proposals for NRSS were primarily focused on collinear AFMs with ${\cal I}$ symmetric crystal structures and zero net magnetization but broken ${\cal T}$ symmetry. These systems exhibit spin splitting that is \emph{even} in momentum, i.e., the splitting energy satisfies \(\Delta \epsilon(\vec{k}) = \Delta \epsilon(-\vec{k})\) and are now commonly referred to as \emph{altermagnets} \cite{Libor2022PRX}. As proposed in Ref.~\cite{Libor2022PRX}, the following identification rules are used to classify such systems:

\begin{itemize}
    \item[(a)] The magnetic unit cell must be identical to the structural unit cell, i.e., the propagation vector \(\vec{k} = 0\), to ensure broken \(\cal T\) symmetry .
    
    \item[(b)] The opposing magnetic sublattices must not be related by inversion symmetry, but rather by certain point group operations such as rotation, or combinations of rotation and translation or inversion. For instance, magnetic atoms connected by a rotation operation, as shown in Fig. \ref{fig3}a, will host spin magnetic moments of equal magnitude. Consequently, their opposite orientations in AFMs would lead to a zero net magnetization within the unit cell.
\end{itemize}

Moving on to the characteristics features in the band structure, the presence of \(\cal I\) symmetry enforces that the spin splitting energy \(\Delta \epsilon(\vec{k})\) is symmetric in momentum space, as dictated by Eq.~(\ref{ITsym}) (see Fig. \ref{fig1}d). This contrasts with SOC-driven Rashba-like spin splitting, which is antisymmetric and associated with continuously varying spin textures in momentum space, as depicted in Fig. \ref{fig1}c. Further to maintain zero net magnetization, \(\Delta \epsilon(\vec{k})\) must change sign along different momentum directions such that the BZ integral of \(\Delta \epsilon(\vec{k})\) over all occupied states vanishes. This requirement results in the formation of \emph{nodal planes}, where spin-up and spin-down bands are degenerate, as shown in Fig. \ref{fig4}. The number of nodal planes intersecting the \(\Gamma\)-point determines the classification of the spin splitting pattern: \(d\)-, \(g\)-, and \(i\)-wave with 2, 4, and 6 nodal planes, respectively \cite{Libor2022}. Moreover, based on their dependence on the out-of-plane momentum component \(k_z\), these can be further classified as: planar or bulk type, as schematically illustrated in Fig. \ref{fig4}a and b \cite{Libor2022}. While the former describes a spin splitting energy independent of \(k_z\), i.e., the spin splitting remains identical across all \(k_z\) planes, for the latter, it is dependent on \(k_z\), resulting in a spin splitting that varies throughout the three-dimensional BZ. While planar spin-momentum locking is relevant for both (quasi)two-dimensional and three-dimensional materials, bulk spin-momentum locking is unique to three-dimensional crystals.}
 
 \begin{figure}[t]
    \centering
    \includegraphics[width=\columnwidth]{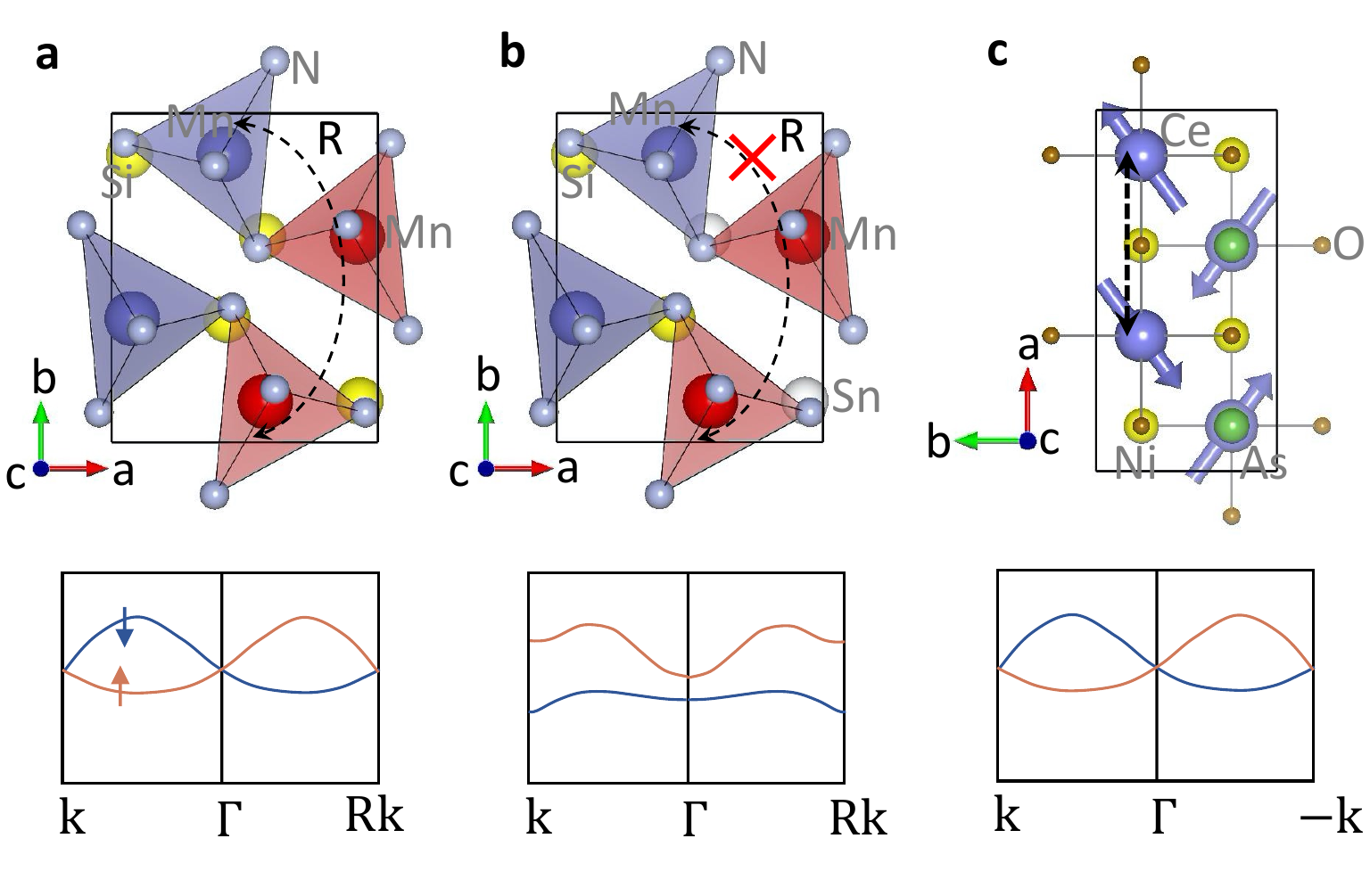}
    \caption{Schematic illustration of (a) symmetric NRSS, (b) NRSS at the BZ-center, and (c) antisymmetric NRSS. The top panel shows the crystal structures of three representative materials, MnSiN$_2$, MnSiSnN$_2$, and CeNiAsO, respectively. While the two magnetic sublattices (marked in blue and red) are connected by a rotational symmetry $R$ in (a), it is absent in (b). The noncollinear coplanar AFM arrangement in (c), on the other hand, has two oppositely oriented spin moments connected by time-reversal plus translation as indicated by the dashed arrow, thereby preserving the global $\cal T$ symmetry. The corresponding band structure features are illustrated schematically in the bottom panel.}
    \label{fig3}
\end{figure}

  {\item {\bf Magnetic unit cell larger than structural unit cell}- Although early studies focused on systems where the magnetic unit cell coincides with the structural unit cell \cite{Libor2022PRX}, this condition is not strictly necessary. The ${\cal T}$ symmetry can also be broken in materials where the magnetic unit cell is larger than the structural one, corresponding to a non-zero magnetic propagation vector $\vec k \ne 0$. This extended class of systems has been explored using first-principles methods, and NRSS has been theoretically demonstrated in such cases \cite{Rodrigo2024}. Similar to the previous case of $\vec k= 0$, these materials can also exhibit $d$-, $g$-, and $i$-wave spin splitting patterns. Candidate materials satisfying these criteria have been identified using ab-initio calculations, further expanding the landscape of potential NRSS systems.}

\begin{figure}[t]
    \centering
\includegraphics[width=\columnwidth]{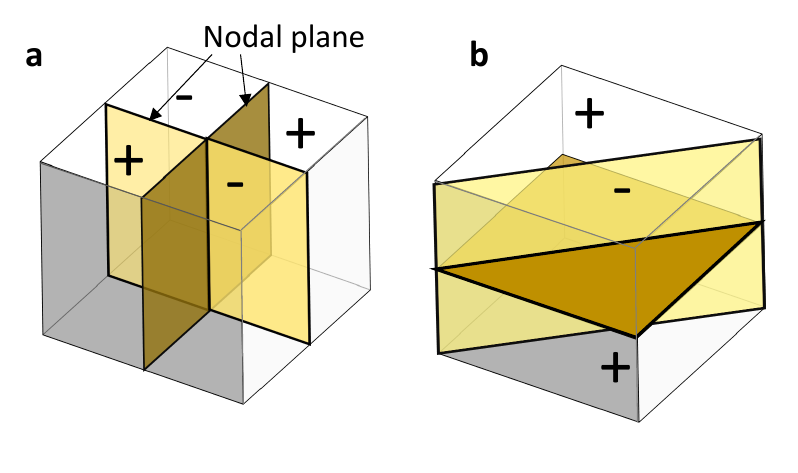}
    \caption{Planar vs. Bulk NRSS. Schematic illustration of (a) planar $d$-wave spin splitting, depicting $k_z$ (vertical direction)–independent spin splitting, and (b) bulk $d$-wave spin splitting, showing the $k_z$ dependence of the NRSS. The planes denote the nodal planes of degenerate up- and down-spin–polarized bands. The $\pm$ sign represents the sign of the energy difference, $\Delta \epsilon(\vec{k})= \epsilon(\vec{k} \uparrow)- \epsilon(\vec{k} \downarrow)$.}
    \label{fig4}
\end{figure}

{\item {\bf Noncollinear AFMs}- 
Beyond collinear antiferromagnetic ordering, noncollinear AFMs can also host NRSS. Notably, NRSS was identified in noncollinear AFMs back in 1989 \cite{Sticht1989}, predating its recognition in collinear systems, though the context and focus were distinct \cite{Chen2014,Kubler2014,Jakub2017}. Prominent examples include noncollinear coplanar AFMs with inversion (${\cal I}$) symmetry, such as the Mn$_3$X Heusler compounds \cite{Chen2014,Kubler2014,Jakub2017}. 

The presence of ${\cal I}$ symmetry in noncollinear AFMs yields even-order NRSS, analogous to the collinear scenario. However, unlike the momentum-independent spin polarization in collinear AFMs, noncollinear AFMs exhibit a complex spin texture in momentum space, with the spin expectation value varying in both magnitude and direction across the BZ. Moreover, instead of nodal planes typical of collinear AFMs as discussed earlier, the Fermi surface in noncollinear systems reflects the symmetry of the underlying lattice. 
}

\end{itemize}

\subsection{$\cal I$ broken, $\cal T$ preserved} \label{case2}

Noncollinear antiferromagnetic order can also give rise to NRSS when $\cal I$ symmetry is broken while preserving $\cal T$ symmetry \cite{Hayami2020PRBR,Brekke2024,Hellenes2024,Chakraborty2024arXiv,Jungwirth2024}. The resulting band structure depends on whether the spin configuration is coplanar or noncoplanar. In the coplanar case, as illustrated in Fig. \ref{fig3}c, opposite spin moments are related by a two-fold ($C_2$) spin rotation about the axis normal to the plane on which the spin moments are lying, combined with a translation, thus preserving $\cal T$ symmetry.

Unlike the $\cal I$-symmetric case discussed earlier, the presence of $\cal T$ symmetry in this case results in antisymmetric spin splitting (see Fig. \ref{fig3}c), i.e., \(\Delta \epsilon(\vec{k}) = -\Delta \epsilon(-\vec{k})\), as follows from Eq. (\ref{ITsym}). Interestingly, the spin polarization of the bands in momentum space is along the axis perpendicular to the spin plane, in contrast to the collinear $\cal I$-symmetric case, where the spin polarization of the bands is the same as the real-space spin direction.

Noncoplanar, noncentrosymmetric AFMs with $\cal T$ symmetry can also exhibit antisymmetric NRSS similarly to the coplanar case. However, the absence of $C_2$ symmetry in spin space due to non-coplanarity of the spins leads to a momentum-dependent spin texture, reminiscent of SOC-driven spin textures in nonmagnetic systems lacking inversion symmetry.

\subsection{$\cal I$ and $\cal T$ broken}\label{case3}

\begin{itemize}
    
{\item {\bf Collinear AFMs}- Although early studies on collinear AFMs with NRSS focused on $\cal I$-symmetric systems, as discussed in Sec. \ref{case1}, the existence of NRSS in collinear AFMs does not require $\cal I$ symmetry. What is essential is the breaking of combined $\cal IT$ symmetry, whose presence would otherwise enforce band degeneracy. Indeed, several $\cal I$-broken systems have been identified that satisfy the NRSS criteria outlined above and exhibit similar band structure features to their $\cal I$-symmetric counterparts \cite{Nag2024,Yuan2024,Hayami2020PRB}.
}

{\item {\bf NRSS at the BZ center}- In collinear AFMs with NRSS as discussed earlier, a key criterion initially identified was that the two spin sublattices are related by a point group operation, such as a rotation [see rule (b) in Sec. \ref{case1}]. However, a subsequent study has proposed collinear AFMs exhibiting NRSS without any symmetry operation connecting the sublattices, as depicted in Fig. \ref{fig3}b \cite{Yuan2024}. In such cases, the magnetization distribution may differ between sublattices, yet zero net magnetization can still be maintained, provided one spin channel is insulating and the orbital moment is quenched. This complies with Luttinger's theorem \cite{LuttingerWard1960,Luttinger1960}, which enforces integer multiples of the Bohr magneton for stoichiometric systems, allowing full compensation when equal numbers of spin-up and spin-down electrons are present. A key consequence of the absence of point group symmetry connecting the sublattices is that NRSS can persist even at the $\Gamma$ point (see Fig. \ref{fig3}b lower panel), in contrast to the symmetry-related sublattices where the rotational operation enforces spin degeneracy at the $\Gamma$ point of the BZ (see Fig. \ref{fig3}a lower panel).}

{\item  {\bf Noncollinear AFMs}- Noncollinear AFMs that break $\cal I$, $\cal T$, and combined $\cal IT$ symmetries can host NRSS of both even and odd order in $\vec k$. In systems with coplanar spin configurations, the resulting NRSS exhibits spin polarization perpendicular to the spin plane, consistent with the behavior discussed in Section~\ref{case2} \cite{Hayami2020PRB}. More recently, Radaelli and Gurung \cite{Radaelli2025} have extended the framework of even-order NRSS to general noncollinear magnets, where it manifests as a momentum-dependent spin texture in reciprocal space. }

\end{itemize}

\section{Candidate Materials: Theoretical Proposals and Experimental Detection} \label{sec3-exp}

Theoretical investigations, particularly symmetry analysis and first-principles calculations, have played a central role in identifying candidate materials that can host NRSS. Interestingly, several well-known antiferromagnetic compounds have recently been recognized to exhibit NRSS. These materials are not inherently rare; rather, the subtle nature of the spin splitting, often manifesting along unusual momentum directions, has delayed their identification. Theoretical proposals of NRSS span a wide range of material systems (see Table~\ref{tab1}), including metals, semiconductors, and insulators.

\begin{figure}[t]
    \centering
     \includegraphics[width=\columnwidth]{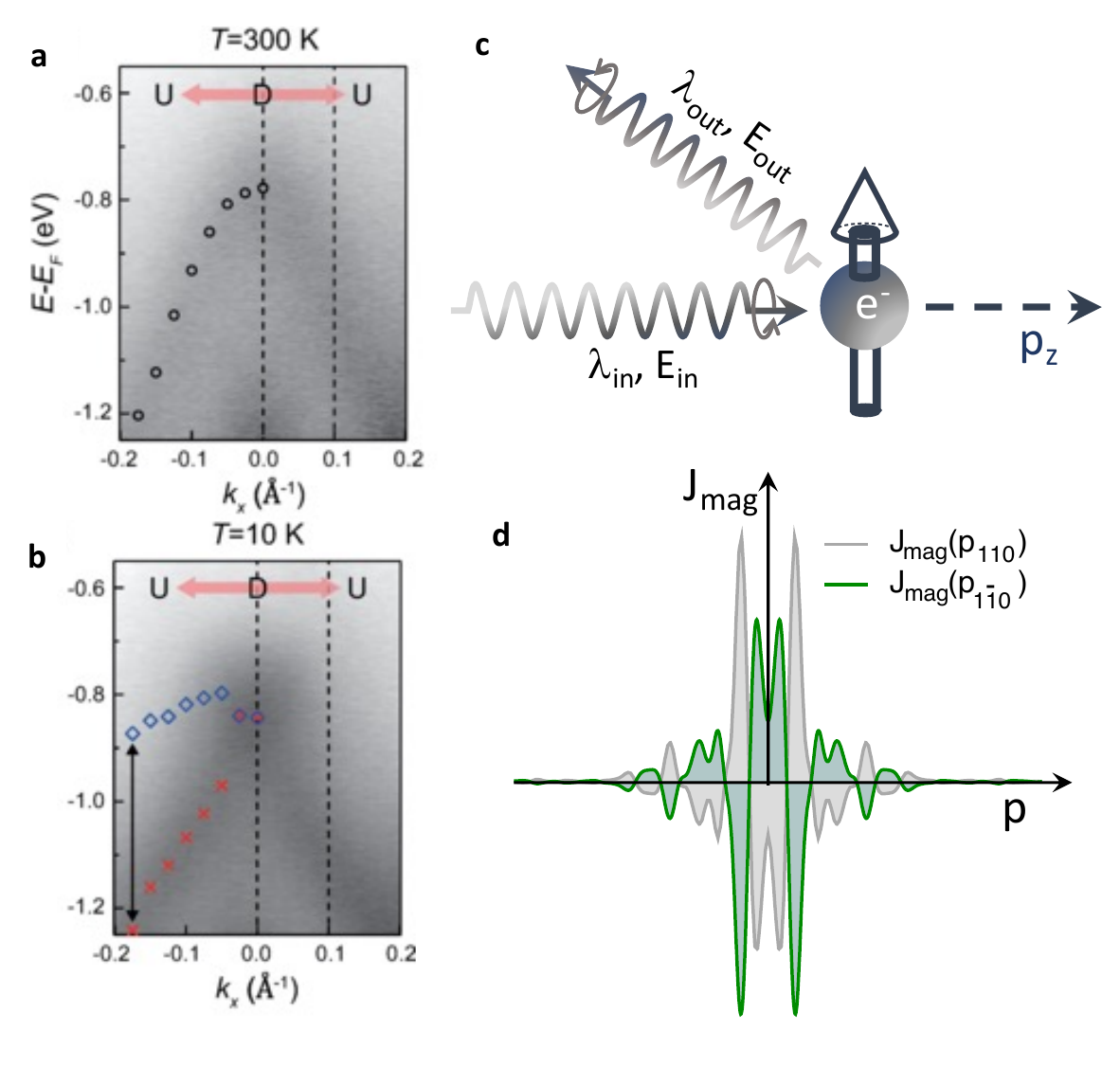}
    \caption{Detection of NRSS. Measured ARPES data along the $D$–$U$ line at (a) $T > T_N$ and (b) $T < T_N$ for $\alpha$-MnTe, as reported in Ref. \cite{Lee2024}. Schematic illustration of (c) magnetic Compton scattering with circularly polarized photons, a photon-in–photon-out process, and (d) the theoretically predicted magnetic Compton profile $J_{\rm mag}(\vec{p})$ in $d$-wave NRSS AFMs, depicting the reversal of the profile as the momentum direction changes from $[110]$ to $[1\bar{1}0]$. Panel (a) and (b) are reprinted with permission from Reference \citenum{Lee2024} Copyright 2024 by the
American Physical Society.}
    \label{fig5}
\end{figure}

\begin{table*}
\caption{Selected representative candidate materials theoretically proposed for various types of NRSS (as discussed in Section \ref{sec2}), along with their structural space groups and electronic conduction properties.}
\label{tab1}
\begin{center}
%\footnotesize
\setlength{\tabcolsep}{7pt}
\begin{tabular}{c c c c c}
\hline
\hline
%& &  & RMT values &  \\
 NRSS pattern & Candidate Materials & Space group   & Conduction
& Reference \\
%& & & & \\
\hline
 & MF$_2$ & $P4_2/mnm$ & Insulator &  \cite{Smejkal2022,Yuan2020,BhowalSpaldin2024,Bhowal2025}\\
 & (M= Mn, Fe, Co) &  &  &  \\
$d$-wave  & LaMnO$_3$ & $Pnma$ & Insulator & \cite{Smejkal2022,Bandyopadhyay2025}\\
  & CaCrO$_3$ & $Pnma$ & Metal & \cite{Makoto2021}\\
  & KRu$_4$O$_8$ &  $I4/m$ & Metal & \cite{Smejkal2022PRX} \\ 
%FeF$_2$ & $P4_2/mnm$ & Insulator & $d$-wave & \cite{Smejkal2022,Bhowal2025}\\
%CoF$_2$ & $P4_2/mnm$ & Insulator & $d$-wave & \cite{Smejkal2022,}\\
\hline
Bulk $d$-wave & CuF$_2$ & $P2_1/c$ & Insulator & \cite{Smejkal2022PRX} \\
Bulk $d$-wave  & MnSe$_2$ & $Pa\bar{3}$ & Insulator & \cite{Rodrigo2024}\\
with $\vec k=(0,0,\frac{1}{3})$ & &  & & \\
\hline
$g$-wave & KMnF$_3$ & $I4/mcm$ &  & \cite{Smejkal2022PRX} \\
Bulk $g$-wave & CrSb, MnTe & $P6_3/mmc$ & Metal & \cite{Libor2022PRX}\\
    & Fe$_2$O$_3$ & $R\bar{3}c$ & Insulator & \cite{Libor2022PRX,Verbeek2024}\\ 
Bulk $g$-wave  & CsCoCl$_3$, RbCoBr$_3$ &  & Metal& \cite{Rodrigo2024} \\
with $\vec k=(\frac{1}{3},\frac{1}{3},0)$ & BaMnO$_3$ & $P6_3mmc$ & Insulator & \cite{Rodrigo2024} \\
\hline
$i$-wave & MnP(S,Se)$_3$ & $P31m$ & Insulator & \cite{Mazin2023}\\
\hline
 & CeNiAsO & $P4/nmm$ & Metal & \cite{Hellenes2024}\\
Anti-symmetric SS & Ba$_3$MnNb$_2$O$_9$ & $P\bar{3}m1$ & Insulator & \cite{Hayami2020PRBR}\\
& FePO$_4$ & $Pnma$ & Insulator & \cite{Hayami2020PRB}\\
\hline
$\Gamma$ point SS & Mn$_2$SiSnN$_4$ & $Pc$ & Semiconductor & \cite{Yuan2024}\\
\hline
$k$-dependent  & Mn$_3$X (X=Sn, Ge) & $P6_3/mmc$ & Metal & \cite{Kubler2014}\\
spin texture &  Mn$_3$GaN &  $Pm\bar{3}m$  & Metal & \cite{Radaelli2025}\\
 & Pb$_2$MnO$_4$ & $P\bar{4}2_1c$ & Insulator & \cite{Radaelli2025}\\
\hline\hline
\end{tabular}
\end{center}
\end{table*}

Following these several theoretical proposals, experimental efforts have been undertaken to detect the NRSS directly. Angle-resolved photoemission spectroscopy (ARPES) is a key tool for probing NRSS in the electronic band structure. Using this technique, a large spin splitting has been confirmed in several candidate materials, including MnTe, CrSb, and CoNb$_4$Se$_8$~\cite{Osumi2024,Wilkinson2024,Dale2024}, consistent with theoretical predictions. However, the presence of NRSS in RuO$_2$, one of the prime candidate materials proposed theoretically, remains under debate \cite{Liu2024RuO2,Zhu2019,Mukuda1999,Hiraishi2024}. Recent experiments show the absence of antiferromagnetic ordering in RuO$_2$ as it remains paramagnetic in its ground state. Thus, even though RuO$_2$ possesses the correct structural symmetry, the absence of antiferromagnetic order prevents it from exhibiting NRSS \cite{Smolyanyuk2024}.

One of the major experimental challenges in detecting NRSS lies in the limitations of conventional ARPES, which primarily probes surface states. This can be mitigated by using thin films, which provide access to near-bulk electronic properties \cite{Wilkinson2024}. Broken Kramers degeneracy can be identified by comparing ARPES spectra measured above and below the magnetic transition temperature. As shown in Fig. \ref{fig5}a and b, degenerate bands appear in the paramagnetic phase, while spin-split bands emerge in the antiferromagnetically ordered state. However, detecting spin polarization via spin-resolved ARPES requires single-domain samples, which poses an additional challenge. In collinear AFMs with symmetric spin splitting, the NRSS reverses with domain orientation, leading to cancellation in multi-domain samples. In contrast, NRSS in coplanar AFMs (as discussed in Section~\ref{case2}) is robust against domain reversal, providing a potential experimental advantage.

While ARPES is an effective tool for probing metallic and semiconducting systems, probing NRSS in insulating AFMs using ARPES is unfeasible. In this context, magnetic Compton scattering (MCS) has been proposed as a promising technique for such materials \cite{BhowalSpaldin2024}. A schematic of MCS is shown in Fig. \ref{fig5}c. MCS measures the spin-polarized electron momentum density via the magnetic Compton profile $J_{\rm mag}(p_z)$ \cite{PlatzmanTzoar1970}:
\begin{equation} \label{jmag}
J_{\rm mag}(p_z) = \int \int \left[\rho_{\uparrow}(\vec p) - \rho_{\downarrow}(\vec p)\right] dp_x\, dp_y,
\end{equation}
a method traditionally applied to ferromagnets and ferrimagnets \cite{SakaiOno1976,Cooper1991,Duffy2013}. In systems exhibiting NRSS without net magnetization, theoretical studies predict two key features in $J_{\rm mag}$ (see \ref{fig5}d): (1) its integral over $p_z$ vanishes, and (2) the profile changes sign under certain rotation of the momentum directions (e.g., ${\cal C}_4$ rotation for $d$-wave symmetry). Although experimental confirmation is still pending, MCS may face challenges similar to ARPES, including the requirement for single-domain samples.

\begin{figure*}[t]
    \centering
\includegraphics[width=\textwidth]{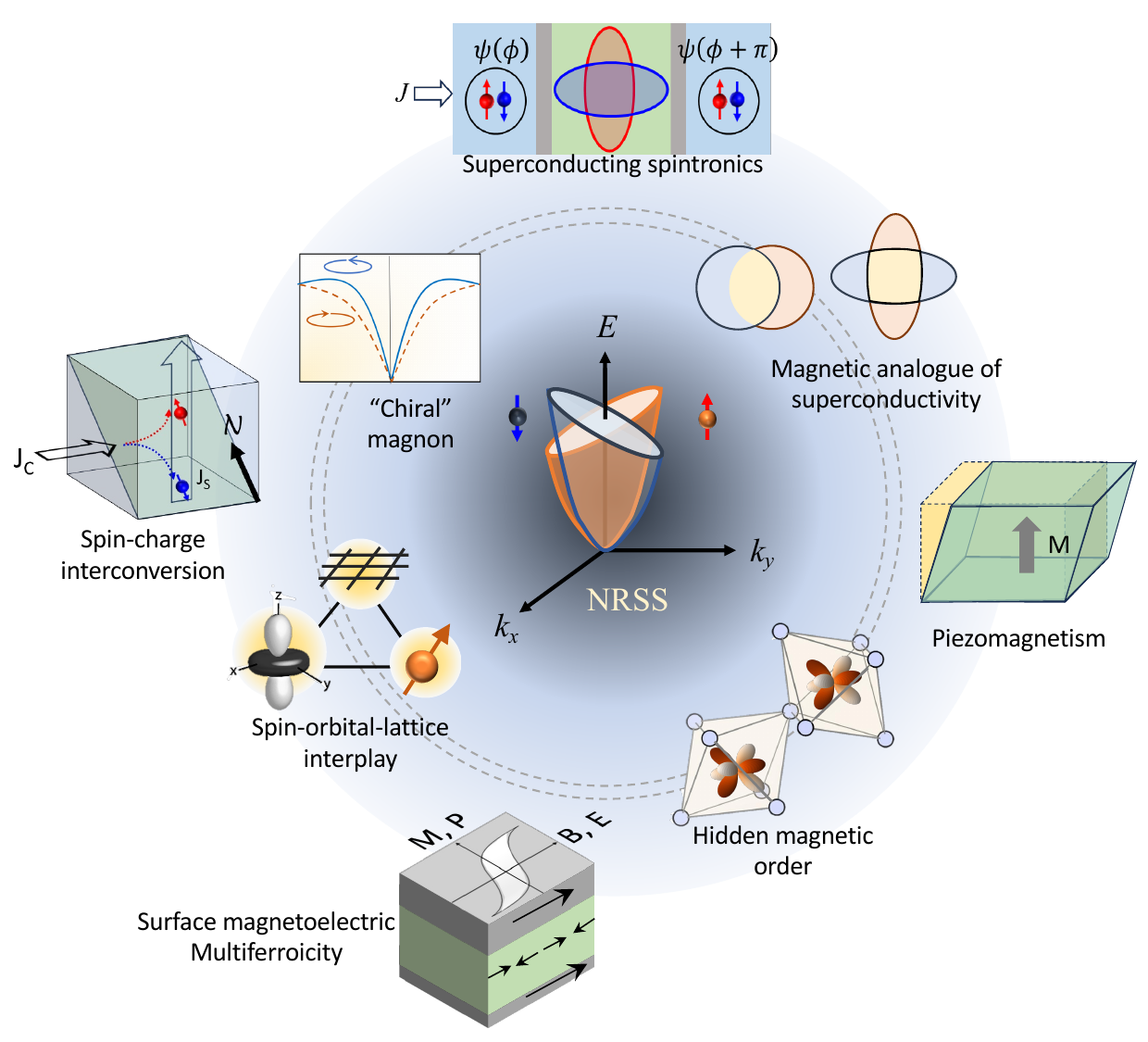}
    \caption{Schematic illustration of emerging concepts (inner circle) and functionalities (outer ring) associated with NRSS.}
    \label{fig6}
\end{figure*}

\section{Advanced Properties and Functionalities}

\begin{figure*}[t]
    \centering
    \includegraphics[width=\textwidth]{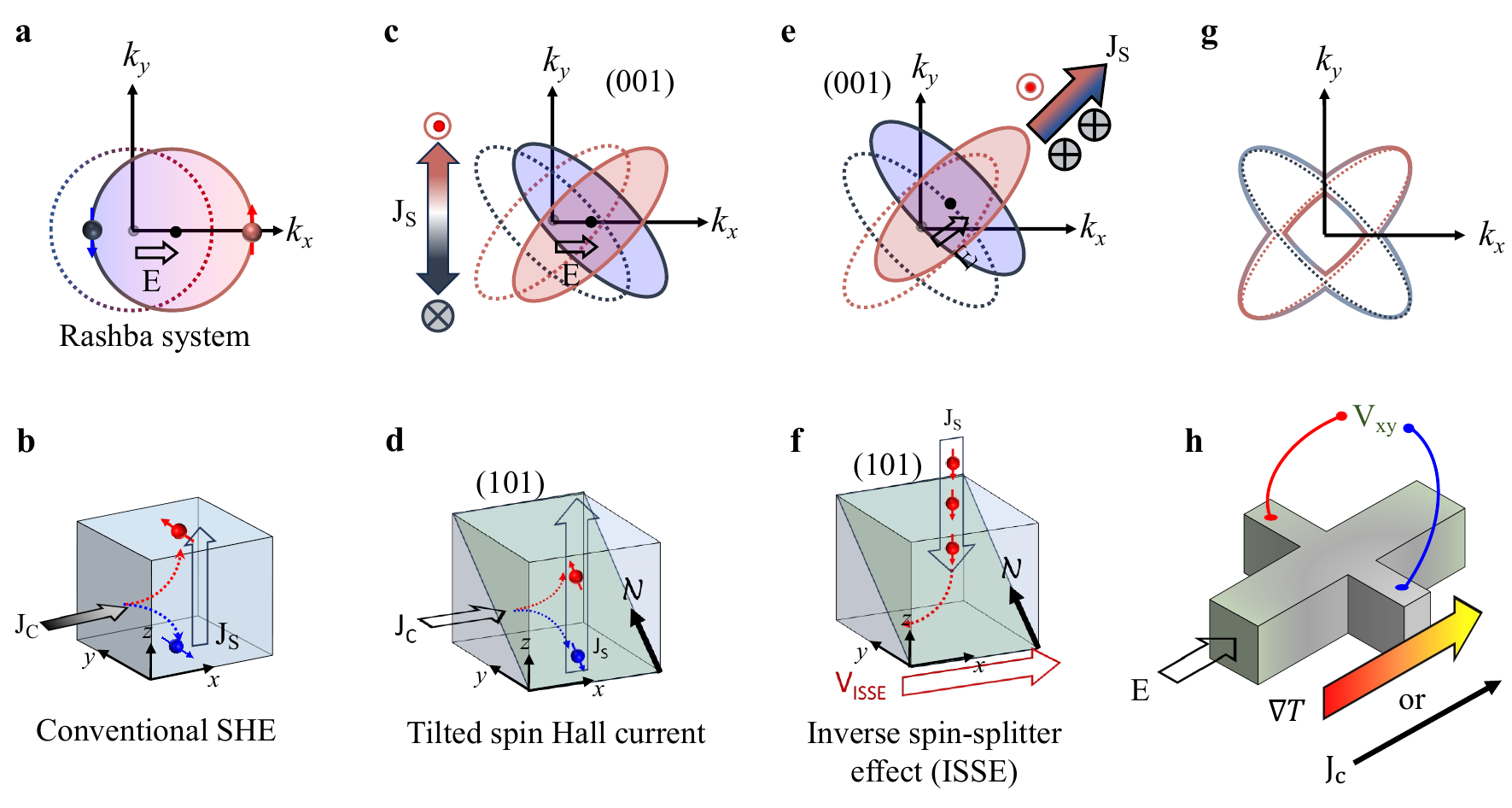}
    \caption{Spin currents from relativistic and NRSS bands. (a) Schematic showing reciprocal-space spin-momentum locking in the relativistic Rashba system at equilibrium (dashed circle) and non-equilibrium (solid circle) conditions. At non-equilibrium, the displacement of the Fermi surface, denoted by the black arrow, leads to the generation of a spin accumulation at the surface. (b) Schematics of transverse spin current $J_s \equiv J_{xz}^y$ generated by the applied charge current $J_c$ due to the spin Hall effect (SHE) driven by SOC. (c) Illustration of a typical NRSS, generating transverse spin current without SOC upon application of an electric field along $\hat x$. (d) The spin polarization, $\sigma$, of the generated $J_s$ in NRSS materials is collinear with the N\'eel vector, $\cal N$,  and depends on the crystal axis, offering a re-orientable spin current, $J_{xz}^\sigma$, unlike the SHE. (e) NRSS producing longitudinal spin-polarised current, $J_{xx}^\sigma$, upon application of an electric field along $[110]$ direction. 
     (f) Spin to charge conversion from the NRSS bands in the inverse spin-splitter effect. (g) NRSS bands together with SOC results in avoided crossing and there by AHE. (h) Schematic experimental set-up for probing AHE and ANE from NRSS effects by applying electric current ($J_C$) and thermal gradient ($\vec \nabla T$), respectively. The color map in (a,c,e,g) schematically represents the orientation of the spins in the $k$ space with the up (down) spin moment denoted by the red (blue) color.}
    \label{fig7}
\end{figure*}

We now turn to key properties and functionalities associated with NRSS. Early interest in NRSS centered on unconventional spin-polarized transport, but research has since expanded to include multiferroicity, chiral magnons, superconductivity, valleytronics in 2D systems, etc. While a comprehensive discussion is beyond the scope of the present review, we highlight a few particularly promising directions in this section.

\subsection{NRSS Effects on Transport Properties}

By combining the contrasting advantages of ferromagnets and AFMs, NRSS bands exhibit several intriguing transport phenomena that are forbidden in $\cal IT$-symmetric AFMs, as we proceed to discuss here. 

\subsubsection{Novel Transverse Spin Current in the Absence of SOC} 

  Traditionally, SOC has been regarded as the key driver of large spin Hall currents, as in heavy metals, topological insulators, Rashba and Dresselhaus systems, where spin-momentum locking (see Fig.~\ref{fig7}a) generates transverse spin accumulation with mutually perpendicular spin polarization, spin flow, and applied current (see Fig.~\ref{fig7}b), facilitating manipulation of adjacent nanomagnets via spin-orbit torques (SOTs) for nonvolatile memory applications \cite{Manchon_2019_Rev}. In contrast, NRSS, arising from broken ${\cal IT}$ symmetry, induces transverse spin currents without SOC under nonequilibrium conditions (see Fig.~\ref{fig7}c,d) \cite{Hernandez2021}. Unlike meV-scale SOC splittings, NRSS can reach $\sim$1 eV, resulting in large spin currents with a predicted spin Hall angle as high as $\sim$34$^\circ$, far exceeding heavy metals \cite{Hernandez2021}.

An important characteristic of the generated spin current is its strong dependence on the crystal axis and the N\'eel vector ($\vec {\cal N}$), providing an additional control knob \cite{Hernandez2021}. Unlike SOC-driven effects, which are even under ${\cal T}$, NRSS-induced spin current reverses sign upon 180$^\circ$ rotation of $\vec {\cal N}$, i.e., odd under $\cal T$. In this context, an important finding is the observation of crystal-axis-dependent tilted spin Hall currents with the spin polarization nearly collinear with $\vec {\cal N}$, as shown in Fig.~\ref{fig7}d.  \cite{Bose2022, Bai2022, Karube2021}, leading to unconventional $z$-polarized spins flowing along $\hat z$. 
This effect, which is symmetry-forbidden in conventional spin Hall systems, allows ``field-free switching" of perpendicular magnets \cite{Karube2021, Fan_2024_ACSNano, Bai_2025_PRApp_MTJ_switching}.

By Onsager reciprocity, the inverse effect of spin-to-charge conversion (see Fig. \ref{fig7}f), known as inverse spin-splitter effect (ISSE), is also possible, allowing detection of out-of-plane polarized spins in spin Seebeck and spin pumping setups using NRSS bands \cite{Bai_2023PRL_ISSE, Lio_2024PRL_ISSE_vs_ISHE_via_SSE, Wang_2024PRLISSE_via_SP}, which conventional ISHE cannot access.

\subsubsection{Longitudinal Spin-Polarized Current with Zero Net Magnetization} 

In ferromagnets, longitudinal spin-polarized currents arise from Zeeman spin-splitting, whereas Kramers degeneracy in compensated AFMs usually forbids such behavior. Interestingly, several NRSS candidates exhibit band structures that permit longitudinal spin polarization despite zero net magnetization, when current flows along specific crystal axes as indicated in Fig.~\ref{fig7}(e,f) \cite{Libor2022TM}. This opens the door to realizing giant magnetoresistance (GMR) and tunneling magnetoresistance (TMR), with predictions that even spin-neutral NRSS bands can yield TMR \cite{Shao2021}. Although AFMs offer benefits such as negligible stray fields and THz operation, their adoption has been hindered by the challenge of electrically reading AFM states. Longitudinal spin-polarized currents overcome this bottleneck, facilitating efficient electrical detection via GMR/TMR and control of AFM bits through spin-transfer torques. Recently, the time-odd nature of this current was also experimentally confirmed \cite{Noh_2025PRL_TMR_RuO2}, marking a key step toward AFM spintronic devices.  

\subsubsection{Anomalous Hall Effect}

The anomalous Hall effect (AHE) is typically observed in FMs, providing an efficient electrical readout of magnetic states \cite{Nagaosa2010}. Conventional AFMs generally lack this effect, but recent studies show that NRSS AFMs, both collinear and noncollinear, can exhibit linear AHE \cite{Chen2014,Libor2020ScAdv,Smejkal2022NatRev}. Noncollinear AFMs, such as Mn$_3$X Heusler compounds, were among the first AFMs identified with AHE \cite{Chen2014}. Later, the discovery of NRSS in collinear AFMs (see Section \ref{sec2}) subsequently expands the pool of AFM materials exhibiting AHE, facilitating AHE-based readout similarly to the longitudinal spin-polarized currents, discussed previously.

Collinear NRSS AFMs exhibit AHE when the N\'eel vector aligns along specific crystal axes under external magnetic fields. Dzyaloshinskii–Moriya interaction, or strain, can also assist in its control. As illustrated in Fig. \ref{fig7}g, avoided band crossings near nodal points in the presence of SOC generate hot-spots of Berry curvature, leading to a large AHE in collinear NRSS AFMs. Alongside AHE, its thermal analogue, the anomalous Nernst effect (see Fig. \ref{fig7}h), has also been observed \cite{Badura2025}.

Since a weak FM moment can be induced for N\'eel vector orientations that allow AHE, it is crucial to isolate contributions from such moments, highlighting the importance of monodomain samples. In this context, probing domain structures in candidate materials, like MnTe, \cite{Amin2024} is an important step forward.
Experimental detection \cite{Feng2022,Betancourt2023,Reichlova2024,Han2024ScAdv,Zhou2025} of AHE in various candidate materials over the past few years, including demonstration of 180$^\circ$ N\'eel vector switching \cite{Han2024ScAdv} and crystal-axis control \cite{Zhou2025}, establishes NRSS AFMs as a new frontier for AFM spintronics.

\subsubsection{Transport through altermagent/superconductor junctions}

\begin{figure}[t]
    \centering
\includegraphics[width=\columnwidth]{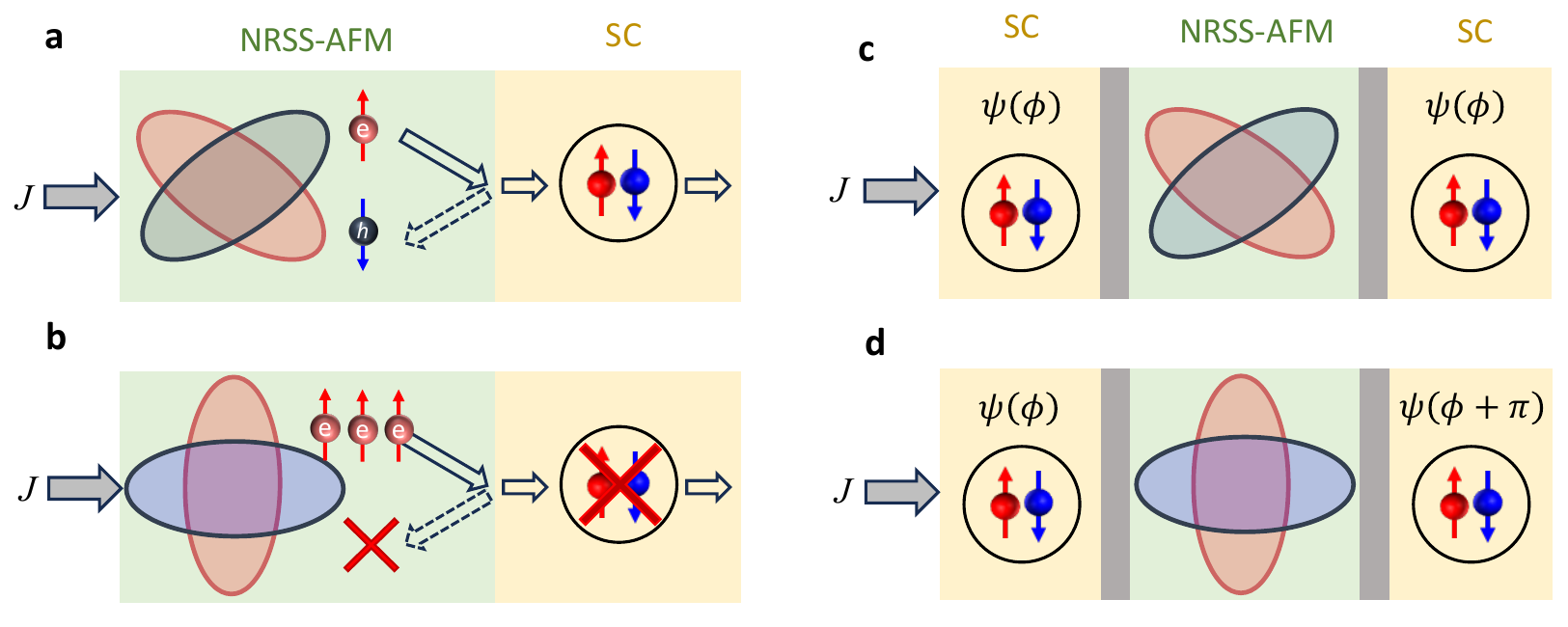}
    \caption{Superconducting spintronics at the junction of a superconductor (SC) and NRSS AFMs. Schematic illustration of (a–b) anisotropic Andreev reflection and (c–d) Josephson effects, controlled by the crystal axis and magnetic ordering (see text for details).}
    \label{fig8}
\end{figure}

The NRSS in AFMs has a conceptual analogue in superconductors (SC). The symmetry of the spin-splitting energy corresponds to the superconducting gap ($\Delta_{\mathrm{sc}}$), which can be isotropic or anisotropic depending on the pairing symmetry, and dictates the phase of the Cooper-pair wavefunction in momentum space. For example, isotropic $\Delta_{\mathrm{sc}}$ in $s$-wave SC parallels isotropic exchange splitting in conventional FMs (Fig.~\ref{fig1}a), while the antisymmetric $p$-wave and symmetric $d$-wave NRSS discussed in Sec.~\ref{sec2} are analogous to unconventional SCs, where $\Delta_{\mathrm{sc}}$ exhibits either a continuous $2\pi$ phase winding (e.g., $\Delta_{\mathrm{sc}} \propto k_x \pm i k_y$) or a $\pi$-phase shift (e.g., $\Delta_{\mathrm{sc}} \propto k_x^2 - k_y^2$), respectively.

Beyond this analogy, SC/NRSS AFM junctions have recently attracted significant attention due to their potential to host phenomena such as Andreev reflection (AR) \cite{Sun2023,Beenakker2023,Papaj2023}, Josephson effects \cite{Ouassou2023,Lu2024}, superconducting diode behavior \cite{Cheng2024,Banerjee2024}, and even Majorana fermions \cite{Ghorashi2024,Nag2025}. In conventional SC/NM junctions, AR converts an incident electron into a reflected hole while transmitting a Cooper pair into the SC, as illustrated in Fig. \ref{fig8}a. For $s$-wave SCs, spin-singlet pairing enforces spin conservation and enhances interface conductance. Consequently, in SC/FM junctions, AR is suppressed by spin polarization. In contrast, the interplay between $s$-wave SCs and $d$-wave NRSS systems allows an anisotropic longitudinal spin current $J_{xx}^s$, which is finite along [110] but vanishes along [100], mimicking FM-like and AFM-like responses, respectively. This anisotropy yields enhanced AR along [100] ($J_{xx}^s \approx 0$) (see Fig. \ref{fig8}a) and suppressed AR (see Fig. \ref{fig8}b) with spin-polarized pairing along [110] ($J_{xx}^s \neq 0$). Such junctions can even separate BCS and spin-polarized Cooper pairs transversely \cite{Giil2024}.

Josephson junctions (JJs) comprising SC/$d$-wave NRSS interfaces further exploit this anisotropy. %While conventional $s$-wave SC JJs exhibit 0-phase differences, SC/FM/SC junctions can realize $\phi$-shifts. Similarly,%
For example, SC/$d$-wave NRSS JJs result in 0- and $\pi$-phases along [100] and [110], as depicted in Figs. \ref{fig8}c and d,  respectively, with recent work predicting tunable $\phi$ phase via exchange field, barrier thickness, crystal orientation, or doping \cite{Lu2024}. NRSS materials also provide a natural platform for superconducting diodes by generating a FM-like spin polarization without any magnetization. Their intrinsic broken $\mathcal{T}$ symmetry supports diode effects without any external fields \cite{Cheng2024,Banerjee2024}, offering a promising route to scalable, field-free superconducting spintronics.

\subsection{Hidden magnetic order} \label{HiddenOrder}

\begin{figure*}[t]
    \centering
    \includegraphics[width=\textwidth]{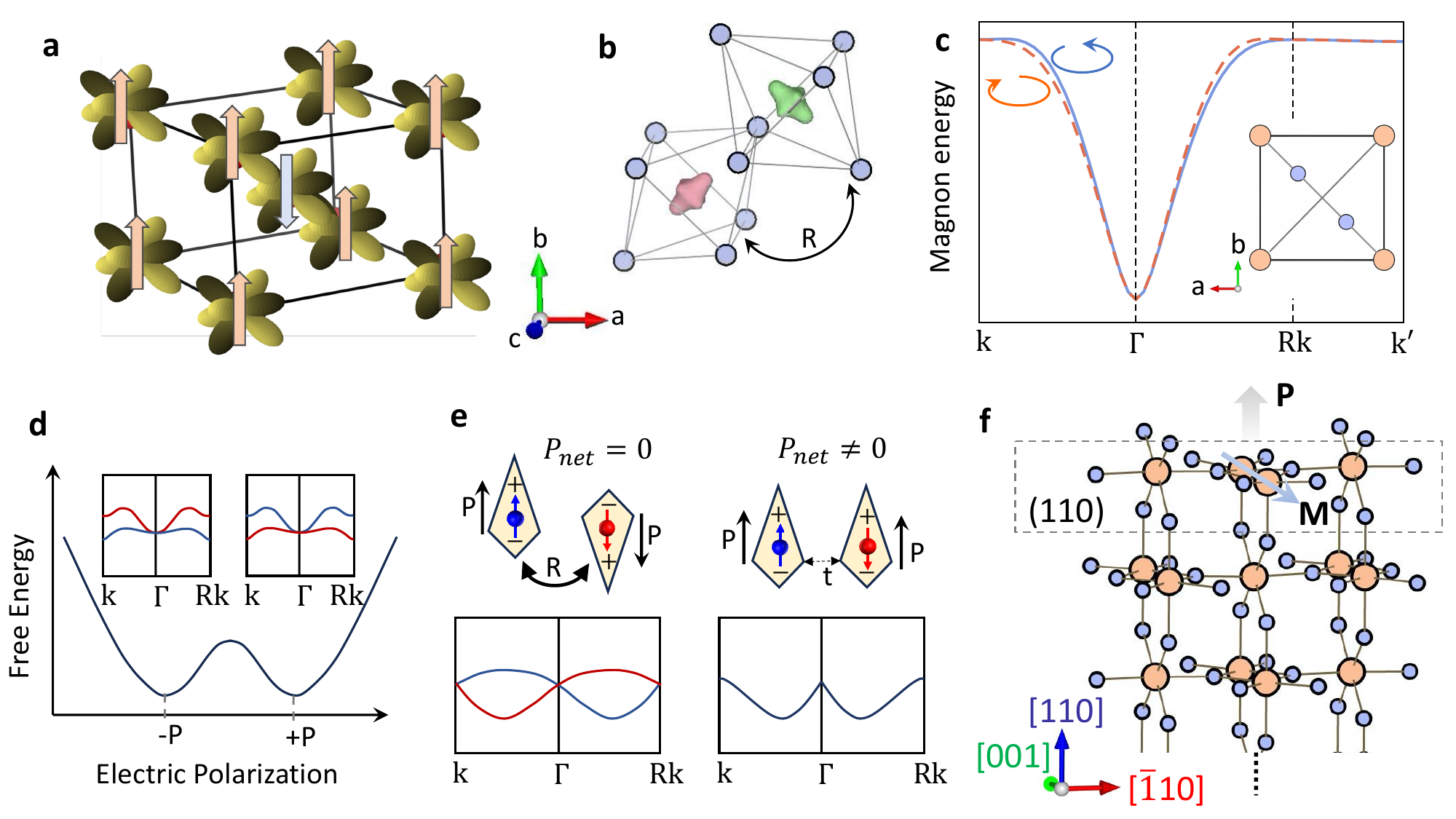}
    \caption{(a) Schematic illustration of hidden ferroically ordered magnetic octupoles, in contrast to antiferroic magnetic dipolar order in $d$-wave NRSS AFMs.
(b) The anisotropic magnetization density at the two magnetic sublattices connected by the rotation $R$, depicting the deviation from the spherically symmetric magnetization density around the magnetic ions and hence indicating the presence of higher-order magnetic multipoles.
(c) The resulting splitting of the magnon bands between magnon modes of opposite handedness. The inset shows the relevant anisotropic exchange interaction responsible for the magnon splitting. Of the two exchange paths, one includes nonmagnetic ligands (shown in blue) while the other does not, leading to differences in the exchange coupling values.
(d) Switching of NRSS with the reversal of ferroelectric polarization $P$. 
(e) Illustration of the switching on ({\it left}) and off ({\it right}) of NRSS during the antiferroelectric ({\it left})-to-ferroelectric ({\it right}) transition.
(f) Schematic illustration of emergent multiferroicity at the (110) surface of the rutile $d$-wave NRSS AFM. 
}
    \label{fig9}
\end{figure*}

Another intriguing aspect of NRSS in compensated AFMs is facilitating the platform for hosting the hidden magnetic order, a topic of long-standing interest \cite{Aeppli2020}. In NRSS AFMs, the broken $\mathcal{T}$ symmetry in the absence of net magnetization indicates the presence of a ferroically ordered magnetic multipole, as illustrated in Fig. \ref{fig9}a, analogous to ferroic magnetic dipole ordering in ferromagnets. Such higher-rank atomic-site magnetic multipoles, e.g., rank-3 magnetic octupoles ${\cal O}_{ijk}=\int \mu_i r_j r_k d\vec{r}$, and rank-5 triakontadipoles $ \tau_{ijklm}=\int \mu_i r_j r_k r_l r_m d\vec{r}$,  describing anisotropic magnetization density beyond the spherically symmetric magnetic dipole (see Fig. \ref{fig9}b), have been proposed in collinear AFMs exhibiting NRSS \cite{Bhowal2022,Verbeek2024}. Here $\vec \mu (\vec r)$ is the magnetization density. Complementary to atomic-site multipoles, cluster and bond multipoles have also been discussed in collinear and noncollinear AFMs with NRSS \cite{Hayami2019,HayamiPRB2020,Suzuki2017,Hayami2020PRB}. Interestingly, as discussed earlier, some of these predictions in noncollinear AFMs predate the identification of NRSS, and were motivated by their role in spin transport \cite{Suzuki2017}.

Interestingly, these higher-order magnetic multipoles were directly linked to NRSS through their $k$-space representations \cite{BhowalSpaldin2024,Verbeek2024,Hayami2019,HayamiPRB2020,Hayami2020}. For $\cal I$ symmetric multipoles, the $k$-space representations can be obtained by substituting $\vec{r} \rightarrow \vec{k}$ in the real-space representations. For example, the magnetic octupole gives rise to a $k$-space representation of form $k_i k_j m_k$, implying NRSS throughout the BZ except on nodal planes $k_i = 0$ and $k_j = 0$ (see Fig. \ref{fig4}a), forming a $d$-wave-like pattern \cite{BhowalSpaldin2024}. Similarly, a $g$-wave NRSS pattern with four nodal planes arises from a rank-5 multipole with fourth-order spatial dependence \cite{Verbeek2024}.

This multipole-based framework provides an alternative to spin group theory for identifying NRSS in unknown systems. Associating specific multipoles with NRSS has far-reaching implications, e.g., it not only helps predict the occurrence of NRSS but also allows tuning its magnitude and sign via controlling the relevant multipole \cite{BhowalSpaldin2024,Verbeek2024,Bandyopadhyay2024, Bandyopadhyay2025}. Furthermore, they can play an important role in identifying related properties of these materials and provide insight into the conjugate fields required to stabilize a particular magnetic domain, crucial for experimental observation, as discussed in Section~\ref{sec3-exp} \cite{BhowalSpaldin2024}.

\subsection{Magnon Splitting} 

Beyond NRSS in the electronic bands of collinear AFMs, recent studies \cite{Libor2023,Bandyopadhyay2024,Hoyer2025} have predicted an analogous splitting in their magnon band structures (see Fig. \ref{fig9}c). This is in contrast to conventional AFMs, where left- and right-handed magnon modes, that are often referred to as ``chiral" magnons, remain degenerate. The emergence of magnon splitting opens up new opportunities for AFM-based magnonic devices without any need for external magnetic fields.

The magnon band splitting has been theoretically proposed for both $d$- and $g$-wave spin-split AFMs \cite{Libor2023, Bandyopadhyay2024, Morano2024, McClarty2024, Hoyer2025,Zhang2025,Wu2025}, and was recently observed in a candidate $g$-wave system \cite{Liu2024}. The symmetry of the magnon splitting reflects that of the electronic spin splitting, reinforcing the correlation between the two. Theoretical studies show that the magnon splitting arises from direction-dependent exchange interactions, which are, in turn, results from the local non-magnetic environment along specific further-neighbor exchange paths, as illustrated at the inset of Fig. \ref{fig9}c. More specifically, the local non-magnetic environment introduces asymmetry in exchange interactions between equidistant bonds, resulting in a stronger coupling along one direction compared to the other, despite having equal interatomic distances. Interestingly, the same non-magnetic environment of the surrounding ligand ions plays a crucial role in forming an anisotropic magnetization density and, consequently, the NRSS in the electronic bands. 

The lifting of degeneracy between magnon modes of specific handedness facilitates selective coupling with polarized neutrons, making polarized neutron scattering a powerful probe for detecting such magnon splitting. 
However, in practice detecting magnon splitting is often challenging, as the energy difference from exchange coupling and hence the magnon splitting energy may lie below experimental resolution \cite{Morano2025}. Nevertheless, given that the magnon and electronic spin splittings share the same symmetry, this technique has been proposed for identifying collinear AFMs exhibiting NRSS \cite{McClarty2024}. Although the magnon splitting is fundamentally non-relativistic in origin, it persists even in the presence of SOC. Notably, in the presence of spin-orbit interaction, it also introduces novel thermal transport effects \cite{Liu2024,Wu2025}.

\subsection{Piezomagnetic effect \& Kinetomagnetism}

NRSS AFMs, although, exhibit zero net magnetization, their intrinsic symmetries allow for the emergence of a net magnetization under certain external perturbations. In this section, we focus on two such effects, the piezomagnetic effect, and kinetomagnetism, through which an external stimulus can induce a net magnetization in otherwise compensated NRSS AFMs.

The piezomagnetic effect describes a change in magnetization, $\Delta M_i$, in response to an applied external stress $\sigma_{jk}$, viz., $\Delta M_i = \Lambda_{ijk} \sigma_{jk}$, where $\Lambda_{ijk}$ denotes the linear piezomagnetic response tensor. 
 The effect can occur in both insulators and metals \cite{Baruchel1988, Disa2020, Ma2021}, both in the absence and presence of SOC \cite{BhowalSpaldin2024}, and is universal to all materials, hosting a ferroically ordered magnetic octupole \cite{BhowalSpaldin2024, Khodas2025, Huyen2025}. Since all $d$-wave spin-split AFMs are ferromagneto-octupolar as discussed in section \ref{HiddenOrder}, they also exhibit linear piezomagnetic effects. However, the effect is not limited to collinear $d$-wave systems and has been predicted for $g$-wave spin-split collinear AFMs as well as for noncollinear AFMs with spin splitting \cite{Khodas2025, Huyen2025}. For example, the linear piezomagnetism can be induced by SOC in $g$-wave systems \cite{Aoyama2024, Verbeek2024}, or the response could be nonlinear (second-order) and present without SOC \cite{Ogawa2025}. 
 Physically, SOC-driven Dzyaloshinskii–Moriya interactions and exchange-driven mechanisms have been proposed to explain the piezomagnetic effect with and without SOC.

The linear piezomagnetic response facilitates tuning both the magnitude and direction of induced magnetization by switching between tensile and compressive stress. Its anisotropic nature in noncollinear AFMs also opens avenues for magnetic device applications \cite{Huyen2025}. Advances in optically induced strain further overcome the limitations of mechanical strain, supporting potential use in memory and spintronic devices \cite{Disa2020, Formisano2022, Formisano2022JPCM}.

Another interesting effect is kinetomagnetism, which describes generation of net magnetization in response to an electric current \cite{Ascher1974}. Unlike the piezomagnetic effect, kinetomagnetism is restricted to metallic systems. Since breaking the combined $\mathcal{IT}$ symmetry is a key requirement for kinetomagnetism to occur, NRSS AFMs that also lack this symmetry are promising candidates to exhibit the effect \cite{cheong2025arXiv}. The symmetry properties of the material, specifically the presence or absence of $\mathcal{I}$, $\mathcal{T}$, and combined $\mathcal{IT}$, dictate whether the induced magnetization is even or odd in the applied current. Kinetomagnetism is closely linked to the AHE. For example, NRSS AFMs with symmetric spin splitting exhibit even-order current-induced magnetization and, consequently, odd-order AHE \cite{cheong2025arXiv}. Conversely, those with antisymmetric spin splitting display odd-order current-induced magnetization and even-order AHE. The current-induced magnetization opens up avenues for electrically controlled magnetization in NRSS AFM systems, further broadening their potential in spintronic applications.

\subsection{Magnetoelectric effect and Multiferroicity}

Similarly to the piezomagnetic effect and kinetomagnetism, many NRSS AFMs exhibit magnetoelectric (ME) effects of various orders, where an external electric field $\varepsilon_k$ induces a net magnetization in otherwise compensated insulating magnets. For example, $d$-wave AFMs hosting magnetic octupoles (see Sec.~\ref{HiddenOrder}) exhibit a second-order ME response, $\Delta M_i = \beta_{ijk} \varepsilon_j \varepsilon_k$, due to a one-to-one correspondence between $\beta_{ijk}$ and the octupole tensor ${\cal O}_{ijk}$ \cite{Urru2022,BhowalSpaldin2024}. Similarly, $g$-wave NRSS AFMs with magnetic triakontadipoles \cite{Verbeek2024} exhibit a fourth-order ME effect. Importantly, however, the leading ME response may differ from that associated with the NRSS-related multipoles. 

A particularly interesting case arises when ME multiferroicity coexists with NRSS \cite{Dong2025}. Here, electric-field control of AFM domains allows reversal of the NRSS spin splitting, $\Delta \epsilon_s = \epsilon_\uparrow - \epsilon_\downarrow$. Similar switching of $\Delta \epsilon_s$ without affecting the N\'eel order has been theoretically shown in improper or geometric ferroelectric \cite{Gu2025,Libor2025}. In such systems, a polar distortion couples to an octahedral rotation mode, connecting the inequivalent AFM sublattices. Reversing the electric polarization reverses the rotation, thereby switching the NRSS, as shown in Fig. \ref{fig9}d. An electric field tuning of NRSS is also predicted in antiferroelectric systems, where a transition to a ferroelectric phase changes the sublattice environment from inequivalent to equivalent, turning NRSS on or off as depicted in Fig. \ref{fig9}e \cite{Duan2025}.

Going beyond the bulk linear ME effect, which requires broken ${\cal I}$ symmetry and is therefore forbidden in ${\cal I}$-symmetric NRSS AFMs, their surfaces can still exhibit this effect \cite{Bhowal2025}.
While $\cal T$ symmetry is already broken in the bulk, $\cal I$ symmetry is broken at the surface, satisfying the symmetry conditions for a linear ME response \cite{Spaldin2013}. Moreover, certain surfaces may exhibit a net electric dipole moment and magnetization, originating from the anisotropy in the bulk magnetization density \cite{Sophie2023, Bhowal2025}, as shown in Fig.~\ref{fig9}f, despite the bulk being a centrosymmetric compensated AFM with zero net moment, leading to the emergence of surface multiferroicity \cite{Bhowal2025}.

\section{Discussions and Future Directions}

The field of NRSS is developing rapidly, and with the inclusion of new types of spin configurations, as discussed in this review, the range of NRSS phenomena under investigation continues to expand. 
Before concluding, we highlight a few emerging directions that we find particularly promising for future research.

Going beyond three-dimensional (3D) bulk materials, NRSS has also been identified in 2D and quasi-2D systems \cite{Zeng2024,Joachim2024,Liu2024PRL}, offering new avenues for exploring spin–valley coupling. In 2D systems, the interplay between NRSS and valley degrees of freedom enables valley polarization \cite{Guo2024PRB,Guo2024PRBL}, offering promising prospects in valleytronics.

With the growing number of material candidates, the functionalities and potential applications of NRSS in antiferromagnetic systems are also expanding. For example, while early studies primarily focused on NRSS-driven spin transport phenomena, recent research has extended the scope to include the magnetic octupolar Hall effect \cite{Han2024}, introducing a new paradigm in spintronics. The detection and clear distinction of this effect from its spin analogue remain open challenges and exciting frontiers for further investigation.

Although this review emphasizes spin splitting arising in the absence of SOC, it is important to recognize that SOC is inevitably present in real materials, even if weak, particularly in light-element systems. A key question, then, is whether NRSS can persist in the presence of SOC. Interestingly, in most known cases, NRSS survives despite the inclusion of SOC. However, SOC can induce additional features, such as SOC-driven spin splitting and topological band crossings \cite{Fernandes2024}. In this context, the recent symmetry-based classification by Cheong and Huang \cite{Cheong2025}, which differentiates between “strong” (NRSS-like) and “weak” (SOC-driven) spin splitting based on magnetic point group symmetries, offers a comprehensive framework to understand spin-split band structures under various conditions. The interplay between NRSS and topological properties is another promising direction for future theoretical and experimental research.

A growing area of interest within the field is the control and tunability of NRSS, especially via external electric fields. Much of this effort has been focused on insulating systems, particularly, (anti-) ferroelectric and multiferroic materials where electric field control offers a path toward low-power spintronic devices. Furthermore, recent theoretical studies proposed a general framework based on phonon-assisted approach suggesting the possibility to control NRSS remains feasible even in inversion symmetric systems through strain, chemical, superlattice engineering, optical excitations, etc \cite{Bandyopadhyay2024,Bandyopadhyay2025}. Detecting NRSS in insulating materials, however, presents technical challenges. In this context, recent experimental efforts of using photovoltaic measurements to probe NRSS are particularly promising \cite{Ezawa2025,Jiang2025,Song2025Nature}, potentially opening a platform for further investigations.

The field of NRSS continues to evolve, fueled by both theoretical innovation and experimental breakthroughs. From broadening the material base to uncovering new physical effects and developing tunable mechanisms, NRSS remains a dynamic and fertile area of research with far-reaching implications in condensed matter physics and next-generation spintronic technologies.

\section*{Acknowledgements}
SB acknowledges many stimulating and informative talks and discussions at the Brainstorming Workshop “Roadmap for Altermagnetism” (NJIT, 2025) and the Fismat Thematic Workshop “Altermagnetism: A New Frontier in Condensed Matter Physics” (2025).
SB gratefully acknowledges financial support from the IRCC Seed Grant (Project Code: RD/0523-IRCCSH0-018), the INSPIRE Research Grant (Project Code: RD/0124-DST0030-002), and the ANRF PMECRG Grant (Project Code: RD/0125-ANRF000-019). AB acknowledges the financial support from IITK seed grant (Project Code: 2023578) and the ANRF PMECRG Grant (Project Code: 2025150). 

\bibliography{reference}

\end{document}